\documentclass[aps,prd,preprint,showpacs,groupedaddress,nofootinbib]{revtex4}
\usepackage{amssymb,amsbsy,color}
\usepackage{latexsym}
\usepackage{graphics}
\usepackage{graphicx}
\usepackage{amsmath}
\usepackage{slashed}
\usepackage{xspace}

\newcommand{\be}{\begin{equation}}
\newcommand{\ee}{\end{equation}}
\newcommand{\bea}{\begin{eqnarray}}
\newcommand{\eea}{\end{eqnarray}}
\newcommand{\ba}{\begin{array}}
\newcommand{\ea}{\end{array}}
\newcommand{\bal}{\begin{align}}
\newcommand{\eali}{\end{align}}
\newcommand{\nn}{\nonumber}

\newcommand{\pref}[1]{(\ref{#1})}

\newcommand{\Sv}{\Sigma_{\rm V}}

\newcommand{\tr}{{\rm tr\,}}

\newcommand{\lt}{\langle}
\newcommand{\rt}{\rangle}

\newcommand{\mpic}{m^2_{\pi^{\pm}}}
\newcommand{\mpin}{m^2_{\pi^0}}
\newcommand{ \dmpi}{\Delta m^2_{\pi}}

\newcommand{\chLog}[1]{\mu_{#1}}
\newcommand{\psibar}{\overline{\psi}}

\begin{document}

\preprint{HU-EP-14/06, SFB-CCP-14-14, TCD-MATH-14-1}
\title{
Charmless chiral perturbation theory for $N_f=2+1+1$ twisted mass lattice QCD
}
\author{$^1$Oliver B\"ar and $^2$Ben H\"orz}

\affiliation{
$^1$Institute of Physics, Humboldt University Berlin,
Newtonstrasse
15, 12489 Berlin, Germany.\\
${^2}$School of Mathematics, Trinity College Dublin, Dublin 2, Ireland.
}

\date{\today}

\begin{abstract}
The chiral Lagrangian describing the low-energy behavior of $N_f=2+1+1$ twisted mass lattice QCD is constructed through O($a^2$). In contrast to existing results the effects of a heavy charm quark are consistently removed. This Lagrangian is used to compute the pion and kaon masses to one loop in a regime where the pion mass splitting is large and taken 
as a leading order effect. In comparison with continuum chiral perturbation theory additional chiral logarithms are present in the results. In particular, chiral logarithms involving the neutral pion mass appear. These predict rather large finite volume corrections in the kaon mass which roughly account for the finite volume effects observed in lattice data.
\end{abstract}

\maketitle

\section{Introduction}
Lattice simulations with twisted mass (tm) Wilson fermions \cite{Frezzotti:2000nk} exhibit various attractive advantages, the most prominent one being automatic O($a$) improvement at maximal twist \cite{Frezzotti:2003ni}. The European twisted mass collaboration (ETMC) has been performing such simulations for a number of years, both with 2 and 2+1+1 dynamical quark flavors.\footnote{For a review of these simulations and the obtained results the reader is referred to Ref.\ \cite{Shindler:2007vp}}
A major disadvantage of twisted mass terms is the explicit breaking of the flavor and parity symmetries, which results in a mass splitting between the charged and neutral pion masses. 

This splitting is a lattice artifact, hence it vanishes in the continuum limit and is not a fundamental concern. 
However, a significant pion mass splitting entangles the chiral and
continuum extrapolation and might lead to non-negligible systematic
uncertainties in taking these limits. Indeed, the splitting is rather large in practice at the lattice spacings simulated. In particular the $N_f=2+1+1$ simulations  show a large pion mass splitting. Table I of Ref.\ \cite{Herdoiza:2013sla} displays the neutral and charged pion masses for seventeen ensembles generated by the ETMC. In seven ensembles the neutral pion mass is less or about equal to 60\% of the charged pion mass, and in only four ensembles the splitting is less than 15\%.  This is a sizable effect, 
and it is expected to modify the way the chiral and
continuum limit is approached.

The impact of the pion mass splitting can be assessed using the appropriate chiral effective theory, so-called tm Wilson ChPT \cite{Sharpe:1998xm,Munster:2003ba,Aoki:2004ta,Sharpe:2004ny}. In the ensembles mentioned before the pion mass splitting is a leading order (LO) effect. The consequences of this power counting have been worked out in Ref. \cite{Bar:2010jk} for the pion masses and the pion decay constant to one-loop order. The main difference to the familiar continuum ChPT results is the presence of chiral logarithms involving both the charged and the neutral pion mass. If the mass splitting is large the chiral extrapolation is influenced in a nontrivial (but calculable) way.

A related source of systematic uncertainties is the finite volume (FV) effects in the simulations. As already pointed out in Ref. \cite{Colangelo:2010cu} the FV effects are substantially larger if the neutral pion mass is smaller than the charged one. A widely used rule of thumb states that FV effects may be ignored if $M_{\pi} L$ is equal to or greater than 4. Even if this rule is satisfied by the charged pion mass it may be violated significantly by the neutral pion mass. Referring again to Table I of Ref. \cite{Herdoiza:2013sla} we find that 14 out of 17 ensembles satisfy $M_{\pi^{\pm}} L \ge 3.8$, while at the same time  only six satisfy  $M_{\pi^{0}} L \ge 3.8$. Three ensembles even have $M_{\pi^{0}} L \le 2$. Since the FV effects are dominated by the smallest particle mass one expects large FV effects, much larger than the estimates based on the charged pion mass.

With these remarks in mind it is natural to ask how observables other than the pion mass and decay constant are affected by a large pion mass splitting. In this paper we give the answer for the simplest observable involving a strange quark, the kaon mass.

Naively one may expect the calculation to be a straightforward extension of the one for the pions in Ref.\ \cite{Bar:2010jk}. This, however, is not the case, for the following reason. 

Twisted mass fermions always come in pairs. The $N_f=2+1+1$ simulations by ETMC involve dynamical strange and charm quarks, in addition to the light up and down type quarks. The standard procedure for the construction of tmWChPT can be applied if both fermions of a pair are either light or heavy. In the first case both fermion flavors give rise to pseudo Goldstone bosons in the chiral effective theory, in the latter they only contribute to the low-energy couplings. For the $N_f=2+1+1$ case this means that the standard procedure to construct tmWChPT treats the $D$ and $D_s$ mesons as pseudo Goldstone bosons just as the pions and kaons. This 4-flavor WChPT has been set up some time ago \cite{Munster:2011gh}. The case of degenerate kaon and $D$-meson masses has been studied even earlier \cite{Abdel-Rehim:2006ve}. However, the results of these papers are valid only for unphysically light charm quarks. They are not applicable to the phenomenologically interesting case with a physical charm quark mass.

For heavy charm quarks the construction of the chiral effective theory needs one additional step. The starting point is the Symanzik effective theory for 2+1+1 flavor lattice theory. Before mapping this 4-flavor theory to ChPT we integrate out charm and construct the effective Symanzik theory for the three light flavors only. This 3-flavor Symanzik theory involves new effective operators for the strange quark, generated by the off-diagonal strange-charm interaction vertices at O($a,a^2$) present in the 4-flavor theory.  These new interaction terms can then be mapped to WChPT using the standard spurion analysis. Depending on the symmetry breaking properties of these additional terms new spurion fields need to be introduced, leading to new terms in the chiral effective theory. The computation of the kaon mass in this effective theory  is then a straightforward extension of the calculation for the pion masses in Ref.\ \cite{Bar:2010jk}.

\section{Twisted mass WChPT without a heavy charm}\label{sect:2}

\subsection{Symanzik effective theory}
The starting point of our analysis is the Symanzik effective action for 4-flavor twisted mass Lattice QCD,
\bea
\label{SYMaction4fl}
S_{\rm Sym}^{(4)} = \int d^4x \left( {\cal L}_0^{(4)} + a {\cal L}_1^{(4)} +a^2 {\cal L}_2^{(4)} +\ldots \right).
\eea
The leading order term ${\cal L}_0^{(4)}$ is the continuum 4-flavor QCD action with the quark mass matrix 
\bea
M^{(4)}=\left(
\begin{array}{cc}
M_l & 0\\
0&M_h
\end{array}\right)\,.
\eea
where $M_{l,h}$ are the $2\times2$ matrices 
\begin{eqnarray}
M_l & = & m + i\mu_l \sigma_3\gamma_5\,,\label{Ml}\\
M_h & = & m + i\mu_h \sigma_1\gamma_5 +\delta \sigma_3\,,\label{Mh}
\end{eqnarray}
and $\sigma^a$ denote the usual Pauli matrices.\footnote{This mass term corresponds to the so-called perpendicular choice according to Ref.\ \cite{AbdelRehim:2006ra}, and it is usually employed by the ETMC in their numerical simulations \cite{Baron:2010bv}. It leads to a real fermion determinant \cite{Frezzotti:2004wz} and has a symmetry that guarantees degenerate masses for all kaons \cite{Chiarappa:2006ae}.} The parts ${\cal L}_{1}^{(4)}$, ${\cal L}_{2}^{(4)}$ capture the cutoff effects of O($a$) and O($a^2$), respectively \cite{Sheikholeslami:1985ij}.

We assume a charm quark mass much heavier than the up, down and strange quark masses. In that case we can integrate out the heavy charm and describe the light flavor physics by an effective 3-flavor theory. In a subsequent step we will map this effective theory into 3-flavor ChPT, the low-energy effective theory involving only the degrees of freedom that are sufficiently light for the chiral expansion.

It is useful to keep this final goal in mind, since it allows substantial simplifications in the following. In order to map into ChPT we are mainly interested in the terms that break chiral symmetry explicitly. The chiral symmetry breaking pattern is mapped into ChPT by the standard procedure called spurion analysis. Nonbreaking terms can be ignored for our purpose. Moreover, different terms in the Symanzik effective theory that break chiral symmetry in the same way are mapped onto the same term in ChPT, so only one representative spurion field for this particular breaking pattern is sufficient \cite{Bar:2003mh}.

It is also useful to keep in mind to which order in the chiral expansion we want to construct the chiral Lagrangian. For our purposes it is sufficient to construct, beside the continuum parts, the terms of O($ap^2, am_q, a^2)$. Including the well-known terms of continuum ChPT through NLO we then have the complete Lagrangian to next-to-leading order (NLO) in the generically small mass (GSM) regime \cite{Sharpe:2004ps}. Also for our main goal, the calculation of the kaon mass to one-loop order in the Aoki or large-cutoff-effects (LCE) regime \cite{Aoki:2003yv,Sharpe:2004ps}  these terms are sufficient. This implies that we can ignore terms in the Symanzik effective action that necessarily generate terms of higher order in the chiral Lagrangian. These include all terms in ${\cal L}_1^{(4)}$ with more than one power of the quark masses $m_q$. Among the terms in ${\cal L}_2$ we need to consider only those without a mass insertion.

In order to remove charm we first rotate from the twisted to the physical basis in the strange-charm sector. In this basis the quark mass matrix assumes the standard diagonal form with real and positive entries. Performing the standard field redefinition (``rotation'')
\begin{eqnarray}\label{fieldrotation}
&&\psi_h \longrightarrow \exp\left(-i\frac{\omega_h}{2}\sigma_1\gamma_5\right)\psi_h\,,\qquad 
\overline{\psi}_h \longrightarrow \overline{\psi}_h \exp\left(-i\frac{\omega_h}{2}\sigma_1\gamma_5\right)\,,
\end{eqnarray}
with 
\begin{equation}\label{cotomegah}
\cot\omega_h = \frac{m}{\mu_h},
\end{equation}
the mass matrix in the heavy sector turns into $M_h ={\rm diag}(m' + \delta, m' -\delta)$, where $m'=\sqrt{m^2 + \mu_h^2}$ denotes the radial mass. Hence we identify the masses for the physical strange and charm quark with $m_s=m'+\delta$ and $m_c=m'-\delta$, respectively.\footnote{We follow the conventions of ETMC and assume $\delta <0$.}

In the physical basis the Lagrangian ${\cal L}_0^{(4)}$ is flavor diagonal. Integrating out charm thus simply amounts to dropping the heavy charm quark part. The quark mass matrix in the effective 3-flavor theory is the $3\times3$-matrix\footnote{For simplicity we drop the superscript in quantities of the 3-flavor theory.}
\begin{equation}\label{M3flavor}
M=
\left(
\begin{array}{cc}
 M_l & 0     \\
 0 &   m_s
\end{array}
\right).
\end{equation}
In ${\cal L}_1^{(4)} $ there is only one term to consider \cite{Sharpe:2004ny}, the Pauli term
\begin{equation}
{\cal L}_1^{(4)} = \Big[\overline{\psi}_l(\sigma F)\psi_l + \overline{\psi}_h(\sigma F)\psi_h\Big].
\end{equation}
Here we introduced the shorthand notation $\sigma F= i\sigma_{\mu\nu}F_{\mu\nu}$. 
Note that we have dropped the unknown coefficient that is multiplying this term. For our purposes this is sufficient. Later on when we map to ChPT the coefficients in the Symanzik effective theory will multiply the low-energy coefficients (LECs) in the chiral Lagrangian. Since both are unknown one usually combines them in single unknown LECs \cite{Bar:2003mh}. 

The Pauli term is flavor diagonal in the twisted basis. In the physical basis there appear off-diagonal terms between the physical charm and strange quark fields,
\begin{eqnarray}\label{Paulired1}
a \overline{\psi}_h(\sigma F)\psi_h &\longrightarrow &a \cos\omega_h \left(\psibar_s \sigma F \psi_s + \psibar_c \sigma F \psi_c\right)
 -i a \sin\omega_h \left(\psibar_s \sigma F \gamma_5 \psi_c + \psibar_c \sigma F \gamma_5 \psi_s \right)\,,
\end{eqnarray}
where the arrow represents the rotation in \pref{fieldrotation}.

Consider first the term proportional to $\cos\omega_h$.
Integrating out charm amounts to dropping the charm field part. 
The remaining strange quark contribution can be combined with the Pauli term for the light flavors and we obtain a Pauli term contribution in the 3-flavor theory that can be conveniently written as
\begin{equation}\label{PauliTerm3f}
a{\cal L}_1 =  \overline{\psi}A(\sigma F)\psi.
\end{equation}
Here $\psi = (\psi_l,\psi_s)$ comprises the light quark field doublet and the strange quark field, and we introduced
\begin{equation}\label{SpurionA}
A= a \, {\rm diag}(1,1, \cos\omega_h)\,. 
\end{equation}
Thinking of the second term in eq.\ \pref{Paulired1} as an interaction  vertex, two of them can be combined, leading to a self-energy diagram with an internal charm quark propagator. Thus, integrating out charm we are left with a new O($a^2$) term involving the strange quark fields only, which is proportional to $a^2\sin^2\omega_h \,\psibar_s (\sigma F)^2 \psi_s$.
There exists no such term involving the light quark fields. Thus, in analogy to \pref{PauliTerm3f} and \pref{SpurionA} we can write
\begin{equation}\label{SquaredPauli}
a^2{\cal L}_{2,1} = \overline{\psi}B(\sigma F)^2\psi,
\end{equation}
where 
\begin{equation}\label{SpurionB}
B= a^2 \sin^2\omega_h 
{\rm diag}(0,0,1)\, \equiv\, a^2 \sin^2\omega_h P_s.
\end{equation}
For later use we have defined the projector $P_s={\rm diag}(0,0,1)$ on the strange quark sector.
Note that this term breaks chiral symmetry like a mass term since $(\sigma F)^2$ commutes with $\gamma_5$.

We also need to check the terms in $a^2 {\cal L}_2^{(4)}$ for new effective operators in the 3-flavor theory. Off-diagonal strange-charm terms are not our concern since at least two of those are needed for a diagonal strange-strange term. Such a term is beyond the order we consider here.

We begin with the 4-quark operators in ${\cal L}_2^{(4)}$. The list of relevant 4-quark operators is given in Ref.~\cite{Bar:2003mh}, eq.~(9), and we follow the notation of this reference.  Consider first the 4-quark term $O_9^{(6)}=(\psibar \psi)^2$ in the 4-flavor theory. Rotating to the physical basis and dropping the off-diagonal parts and an irrelevant 4-charm-quark term we find
 \begin{equation}\label{O9}
a^2 O_9^{(6)} = (\psibar A\psi)^2 +2 a \cos\omega_h (\psibar A\psi) \psibar_c\psi_c - 2a^2 \sin^2\omega_h \psibar_s\gamma_5\psi_c \psibar_c\gamma_5\psi_s\,,
\end{equation}
where we used $A$ defined in \pref{SpurionA}. The first term here is the $O_9^{(6)}$ analogue in the 3-flavor theory. Integrating out the heavy charm quark results in a charm quark propagator in the second and third term, leaving behind fermion bilinears involving the light quark fields only. The third term leads to $a^2 \sin^2\omega_h \psibar_s\psi_s$, which is an O($a^2$) mass term for the strange quark.  It breaks chiral symmetry like the squared Pauli term in \pref{SquaredPauli}, hence both are mapped onto the same term in ChPT and for the spurion analysis in the next section we will need only one of them. The second term in \pref{O9} generates 
\begin{equation}\label{L22}
a^2 {\cal L}_{2,2}= \psibar C \psi\,,
\end{equation}
with 
\begin{equation}\label{SpurionC}
C\,=\,a^2\cos \omega_h {\rm diag}(1,1,\cos\omega_h) = a \cos\omega_h A \,.
\end{equation}
Note that \pref{L22} is a (flavor dependent) mass term of O($a^2$).

Next consider the 4-quark term $O_{10}^{(6)}=(\psibar\gamma_5 \psi)^2$ in the 4-flavor theory. In this case we find
 \begin{equation}\label{O10}
a^2 O_{10}^{(6)} = (\psibar A\gamma_5\psi)^2 +2 a \cos\omega_h (\psibar A\gamma_5\psi) \psibar_c\gamma_5 \psi_c - 2a^2 \sin^2\omega_h \psibar_s\psi_c \psibar_c\psi_s\,.
\end{equation}
Besides the expected $O_{10}^{(6)}$ analogue we obtain again two more terms. However, integrating out charm they generate the same effective operators as $O_{9}^{(6)}$. Therefore, no additional spurion fields need to be introduced later on. 
The same analysis applies to four more 4-quark operators listed in Ref.\ \cite{Bar:2003mh}: $O_{13}^{(6)} - O_{15}^{(6)}$ and $O_{18}^{(6)}$. Again, none of them requires the introduction of additional spurion fields later on. 

The remaining four 4-quark operators are all chiral symmetry preserving and do not lead to additional spurion fields. As a concrete example consider $O_{11}^{(6)} =(\psibar\gamma_{\mu}\psi)^2$. Rotating to the physical basis this reads
 \begin{equation}\label{O11}
 a^2 O_{11}^{(6)} = a^2 (\psibar \gamma_{\mu} \psi)^2 + 2 a^2 (\psibar \gamma_{\mu}\psi) \psibar_c\gamma_{\mu}\psi_c + a^2(\psibar_c \gamma_{\mu} \psi_c)^2\,.
\end{equation}
Integrating out charm the last term can be ignored immediately. 
The first term results in the 3-flavor analogue of $O_{11}^{(6)}$, while the second term leaves behind a new nontrivial term in the effective 3-flavor theory.  The form of this operator is constrained by the symmetries of $a^2 (\psibar \gamma_{\mu}\psi) \psibar_c\gamma_{\mu}\psi_c$. For our purposes here it is sufficient to know that this operator too does not break chiral symmetry. It is therefore irrelevant for the spurion analysis in the next section. 

So far we have discussed the {\em new} operators in the 3-flavor Symanzik theory that are generated by integrating out the heavy charm. We have seen that in all cases the expected 3-flavor 4-quark operators also emerge. Among those are also operators that break chiral symmetry. However, for those the discussion in Ref.~\cite{Bar:2003mh} applies: Taking twice the spurion field introduced at O($a$) is sufficient to generate all terms in the chiral Lagrangian that one would also obtain by introducing separate O($a^2$) spurion fields.

Finally, we turn to the quark bilinears at O($a^2$) in ${\cal L}_2^{(4)}$. Equation (8) in Ref.~\cite{Bar:2003mh} lists eight of them. We need to consider only those without any quark mass insertion, which leaves $O_1^{(6)} - O_4^{(6)}$. These four operators do not break chiral symmetry, which also implies that they are flavor diagonal in the physical basis. For example, $O_1^{(6)}=\psibar \slashed{D}^3 \psi$ reads in the physical basis
\begin{equation}
a^2 O_1^{(6)}= a^2 \psibar \slashed{D}^3 \psi +a^2 \psibar_c \slashed{D}^3 \psi_c\,.
\end{equation}
Integrating out charm we can simply drop the second term. The remaining first part is still chiral symmetry preserving, so no spurion field is needed to map it into ChPT. The same holds for the other three quark bilinears at O($a^2$).

\subsection{Spurion analysis}

The chiral Lagrangian is now constructed following the familiar spurion analysis described in Ref.~\cite{Bar:2003mh}. Each term in the 3-flavor Symanzik effective action that breaks chiral symmetry explicitly is made invariant by introducing a spurion field which transforms nontrivially under chiral transformations such that the whole term is invariant. These spurion fields are then used in the chiral effective theory to write down the most general chiral Lagrangian which is invariant under  chiral symmetry. Once this is achieved the spurion field is set to its original value from the Symanzik effective theory.

In the following we list all the representative spurion fields we need for the construction of the chiral Lagrangian. As explained before, there are many more terms in the Symanzik effective action, but spurion fields that transform the same way and have the same final value will lead to the same term in the chiral Lagrangian. It is therefore enough to consider only one representative field.

At O($a^0$) there is just the mass matrix spurion field which transforms as usual \cite{Bar:2003mh}
\begin{equation}
 M\,\rightarrow LMR^{\dagger} \,,\qquad  M^{\dagger} \,\rightarrow \, RM^{\dagger} L^{\dagger},
\end{equation}
and its final value is given in \pref{M3flavor}. Note that the respective mass matrix $M_l$ reads 
\begin{equation}
M_l =  m + i\mu_l \sigma_3\,,
\end{equation}
i.e.\ without the $\gamma_5$ in the twisted mass contribution. 

In order to make the term in \pref{PauliTerm3f} invariant we promote $A$ to a spurion field that transforms according to 
\begin{equation}
 A\,\rightarrow LAR^{\dagger} \,,\qquad  A^{\dagger} \,\rightarrow \, RA^{\dagger} L^{\dagger}.
\end{equation}
Its final value is given in \pref{SpurionA}.

At O($a^2$) we have found two new operators, cf.~\pref{SquaredPauli} and \pref{L22}. These are made invariant by the  spurion fields $B$ and $C$. Both of them transform in the same way,
\begin{eqnarray}
&& B\,\rightarrow LBR^{\dagger} \,,\qquad  B^{\dagger} \,\rightarrow \, RB^{\dagger} L^{\dagger},\nonumber\\
&& C\,\rightarrow LCR^{\dagger} \,,\qquad  C^{\dagger} \,\rightarrow \, RC^{\dagger} L^{\dagger}\,.
\end{eqnarray}
However, the final values are different, see eqs.~\pref{SpurionB} and \pref{SpurionC}, respectively.

Additional O($a^2$) spurions stemming from terms like the first one in eq.\ \pref{O9} need not be introduced, since these transform the same way as squares of the spurion field $A$, i.e.\ as $A^2,AA^{\dagger}, A^{\dagger}A, (A^{\dagger})^2$, and they have the same final value \cite{Bar:2003mh}. 

\subsection{The chiral Lagrangian}

The chiral Lagrangian can now readily be written down.
The LO continuum part reads (in Euclidean space time) 
\bea\label{LOLagrangian}
{\cal L}_{2} & =& \frac{f^{2}}{4} \lt \partial_{\mu}\Sigma \partial_{\mu}\Sigma^{\dagger}\rt  - \frac{f^{2}}{4} \lt \chi^{\dagger} \Sigma + \Sigma^{\dagger}\chi\rt,
\eea
where $\langle\ldots\rangle$ stands for the trace over flavor indices and the LEC $f$ is the  pseudoscalar
decay constant in the chiral limit.\footnote{Our convention is 
such that $f_\pi=92.21$~MeV.}  The field $\Sigma=\Sigma(x)$ is an element of SU(3) containing the pseudoscalar fields in the usual way,
\bea
\Sigma(x) &=& \Sv^{1/2} \Sigma_{\rm p}(x) \Sv^{1/2},\label{DefSigma}\\[2ex]
\Sigma_{\rm p}(x)&=& \exp \left( \frac{2 i}{f} \sum_{a=1}^{8} \pi^{a}(x) T^{a}\right)\,.
\label{DefSigmaphys}
\eea
$\Sv$ denotes the ground state of the theory, defined as the minimum of the potential energy (density). In continuum ChPT it is simply the identity matrix and has no impact. However, for twisted mass terms $\Sv$ is nontrivial \cite{Munster:2004am}.
The pseudoscalar fields $\pi^{a}(x)$ are real-valued, and we choose the SU(3) group generators to be normalized according to 
$\tr T^{a}T^{b} \, =\, \delta^{ab}/2$.
The subscript `${\rm p}$' in \pref{DefSigmaphys} refers to {\em physical} fields, meaning that the mass terms for the pseudoscalar fields are non-negative and that there are no interaction terms involving less than three pseudoscalars.
The parameter $\chi$ in the chiral Lagrangian  contains the second LO LEC $B_0$ and the mass matrix 
$M$ in \pref{M3flavor} via
\bea
\label{Defchi}
\chi &=& 2B_0M\,.
\eea
Note that the mass matrix is not diagonal and real for twisted mass terms, so $\chi^{\dagger}\neq\chi$ in general. 

The NLO Lagrangian is the one given by Gasser and Leutwyler in Ref.\ \cite{Gasser:1984gg}. We omit a few terms that are not needed for the calculations in this paper: 
\bea
{\cal L}_{{4}}& = & - L_{1}\, \lt \partial_{\mu}\Sigma \partial_{\mu}\Sigma^{\dagger}\rt^{2}\nn\\
& & - L_{2} \,\lt \partial_{\mu}\Sigma \partial_{\nu}\Sigma^{\dagger}\rt \,\lt \partial_{\mu}\Sigma \partial_{\nu}\Sigma^{\dagger}\rt\nn\\
& &-L_{3} \,\lt\partial_{\mu}\Sigma \partial_{\mu}\Sigma^{\dagger}\rt^{2}\nn\\
& & + L_{4} \,\lt \partial_{\mu}\Sigma \partial_{\mu}\Sigma^{\dagger}\rt\, \lt\chi^{\dagger} \Sigma + \Sigma^{\dagger}\chi\rt\nn\\
& & + L_{5}\, \lt\partial_{\mu}\Sigma \partial_{\mu}\Sigma^{\dagger}( \chi^{\dagger} \Sigma + \Sigma^{\dagger} \chi)\big\rt\nn\\
& & - L_{6}\, \lt \chi^{\dagger}\Sigma + \Sigma^{\dagger}\chi\rt^{2}\nn\\
& & - L_{7}\, \lt\chi^{\dagger}\Sigma - \Sigma^{\dagger}\chi\rt^{2}\nn\\
& & -L_{8}\, \lt \chi^{\dagger}\Sigma\chi^{\dagger}\Sigma + \Sigma^{\dagger}\chi \Sigma^{\dagger}\chi\rt.
\eea
The coefficients $L_{i}$ are the well-known Gasser-Leutwyler (GL) coefficients. 

The O($a$) terms in the chiral Lagrangian are obtained with the spurion field $A$. The spurion field $A$ transforms like the mass spurion $M$. Hence, we get the same terms as in ${\cal L}_2$ and ${\cal L}_4$ with $M$ replaced by $A$ \cite{Rupak:2002sm}. The leading term reads 
\bea
\label{lagordera}
{\cal L}_a = - \frac{f^{2}}{4}  \lt \rho (\Sigma + \Sigma^{\dagger})\rt,
\eea
where $\rho=\rho^{\dagger}$ is defined by 
\bea
\rho&=& 2W_0 A\,.
\eea
$W_0$ is a LEC of mass dimension 3, such that $\rho$ has mass dimension 2, just as $\chi$. 
Note that $\rho$ is not flavor diagonal for $\cos\omega_h\neq 1$. Therefore it cannot be taken out of the trace in flavor space. 

At higher order there are terms of O($ap^2,aM,a^2)$ \cite{Rupak:2002sm,Bar:2003mh},
\bea
{\cal L}_{ap^2} & = & \phantom{+}W_{4} \,\lt \partial_{\mu}\Sigma \partial_{\mu}\Sigma^{\dagger}\rt \, \lt \rho(\Sigma + \Sigma^{\dagger})\rt\nn\\
& & + W_{5} \, \lt \partial_{\mu}\Sigma \partial_{\mu}\Sigma^{\dagger} (\rho\Sigma + \Sigma^{\dagger}\rho)\big\rt\,,\nn\\[1ex] 
{\cal L}_{aM} & = &- W_{6}\, \lt\chi^{\dagger}\Sigma + \Sigma^{\dagger} \chi\rt \lt\rho( \Sigma + \Sigma^{\dagger})\rt\nn\\
& & - W_{7} \, \lt\chi^{\dagger}\Sigma - \Sigma^{\dagger} \chi\rt\lt \rho(\Sigma - \Sigma^{\dagger})\rt\nn\\
& & -W_{8} \, \lt \chi^{\dagger} \Sigma\rho\Sigma + \Sigma^{\dagger}\rho\Sigma^{\dagger}\chi\rt,\nn\\[1ex]
{\cal L}^{(1)}_{a^2} & = & - W^{\prime}_{6}\, \lt\rho(\Sigma + \Sigma^{\dagger})\rt^2\nn\\
& & - W^{\prime}_{7} \, \lt\rho(\Sigma - \Sigma^{\dagger})\rt^2\nn\\
& & -W^{\prime}_{8}\, \lt\rho\Sigma\rho\Sigma + \Sigma^{\dagger}\rho\Sigma^{\dagger}\rho\rt.\label{La2}
\eea
The coefficients $W_i, W^{\prime}_i$ are dimensionless LECs, just as the Gasser-Leutwyler coefficients.
The superscript in ${\cal L}^{(1)}_{a^2}$ serves as a reminder that it does not contain all O($a^2$) terms. Additional ones stem from the spurion fields $B$ and $C$, which read
\begin{eqnarray}
{\cal L}^{(2)}_{a^2}&=& -  W_B \hat{a}^2 \sin^2\omega_h \lt P_{\rm s} (\Sigma+\Sigma^{\dagger})\rt\nonumber\\
&& - W_C \hat{a} \cos\omega_h \lt \rho (\Sigma +\Sigma^{\dagger})\rt\label{La22}
\end{eqnarray}
Here we introduced the scaled lattice spacing
\begin{equation}
\hat{a}=2W_0 a\,,
\end{equation}
which has dimension 2 and makes the new LECs $W_x$ in \pref{La22} dimensionless. 

In WChPT it is usually convenient to absorb the leading O($a$) term in \pref{lagordera} in the so-called shifted quark mass \cite{Sharpe:1998xm}. However, here this term is not flavor independent and it depends on $\omega_h$. In order to discuss the dependence of the pion and kaon masses on this angle we prefer to leave this term explicit. We also keep the O($a^2$) terms in \pref{La22} explicit, even though these are mass terms too. 

Finally, we emphasize that all the LECs in the chiral Lagrangian depend on the mass of the heavy charm quark. Therefore, the LECs are constants only if the (physical) charm quark mass is kept fixed.

\section{Pseudoscalar masses in the LCE regime}\label{sec:massesLCE}

\subsection{Preliminaries}
In this section we compute the pseudoscalar masses to one-loop order in the LCE regime \cite{Aoki:2003yv,Sharpe:2004ny,Aoki:2008gy}. This regime assumes that the O($a^2$) cutoff effects are of the same order in the chiral expansion as the effects due to the quark masses. More precisely, it assumes that the O(${a^2}$) terms in the effective Lagrangian contribute to LO,
\bea\label{LOLagLCE}
{\cal L}_{\rm LO} = {\cal L}_2 + {\cal L}^{(1)}_{a^2}+{\cal L}^{(2)}_{a^2}.
\eea
The O($a^2$) terms lead to the mass splitting between the charged and the neutral pion \cite{Scorzato:2004da}. This splitting is a LO effect in the LCE regime counting. Thus, it is the appropriate one if the size of the splitting is observed to be of the same order as the charged pion mass. As discussed in the introduction this is indeed the case for a substantial part of the ETMC data.

The Lagrangian \pref{LOLagLCE} results in interaction vertices of O($a^2$).  These contribute to the pseudoscalar masses at one loop. Our  calculation here follows the one in Ref.\ \cite{Bar:2010jk}, where the pion masses were computed in the 2-flavor theory.  The reader is referred to this reference for all aspects of the calculation that are independent of the number of flavors. In particular, we work at maximal twist only. This simplifies the calculation significantly and it is the relevant case for practical applications.

\subsection{Gap equation and maximal twist}
As a first step we need to compute the ground state $\Sigma_{\rm V}$. The ground state is nontrivial (i.e.\ $\Sigma_{\rm V}\neq1$) since the light sector is still formulated in the twisted basis. On the other hand, we have already rotated the heavy sector into the physical basis, so the nontrivial part of $\Sigma_{\rm V}$ is in the light sector only and the ground state assumes the form \cite{Sharpe:2004ps}
\begin{equation}
\Sigma_{\rm V} = \left(
\begin{array}{cc}
  e^{i\phi_l\sigma_3} &      \\
  & 1      
\end{array}
\right)\,.
\end{equation}
$\phi_l$ is called the (light) vacuum angle. It is determined by minimizing the potential energy density in the chiral Lagrangian. The calculation is essentially as in the 2-flavor case in Ref.\ \cite{Aoki:2004ta}, and we obtain  the following gap equation:
\begin{equation}\label{GE}
2B_0\mu_l \cos\phi_l = \sin\phi_l (2B_0m +2\tilde{W}_0a - 2c_2a^2 \cos\phi_l).
\end{equation}
For better comparison with the 2-flavor result  we have defined the LEC combinations
\begin{equation}
c_2= -32W^{\prime}_{68}\frac{W_0^2}{f^2}\,
\end{equation}
with $W^{\prime}_{68}=2W^{\prime}_6 + W^{\prime}_8$, and 
\begin{equation}
\tilde{W}_0 = W_0\left(1+ \frac{4\hat{a}}{f^2} \cos\omega_h (4W_6^{\prime} + W_C)\right)\,.
\end{equation}
$c_2$ is a well-known LEC in the 2-flavor theory. It determines the phase structure of the theory \cite{Munster:2004am,Sharpe:2004ps} as well as the pion mass splitting at tree-level \cite{Scorzato:2004da}.

The gap equation has the same form as in the 2-flavor theory discussed in \cite{Aoki:2004ta}, the only difference is the explicit appearance of $\tilde{W}_0$ instead of $W_0$ on the right hand side of \pref{GE}. 

The gap equation determines the vacuum angle as a function of the three mass parameters $m,\mu_l,\mu_h$ and the lattice spacing $a$. As has been shown in Ref.\ \cite{Aoki:2004ta} maximal twist and automatic O($a$) improvement is achieved for $\cos\phi_l=0$. The gap equation \pref{GE} immediately tells us that this is a solution only if
\begin{equation}\label{CondMaxtwist}
2B_0m + 2\tilde{W}_0 a=0\,.
\end{equation}
This equation determines the (critical) untwisted mass $m$ to LO. Note that $m$ depends on $\mu_h$ and $a$, but it is independent of $\mu_l$. Note also that $m$ enters \pref{CondMaxtwist} twice, not only explicitly but also implicitly via $\cos\omega_h$. This dependence, however, is expected to be very small since it is an O$(a^2)$ contribution in \pref{CondMaxtwist}.

In practice it is not necessary to satisfy \pref{CondMaxtwist} exactly. It can be shown \cite{Aoki:2004ta,Sharpe:2004ny,Sharpe:2005rq,Aoki:2006nv,Aoki:2006gh} that any mistuning of $m$ that leads to $\cos\phi_l={\rm O}(a)$ is sufficient for automatic O$(a)$ improvement. 

Equation \pref{CondMaxtwist} can be solved iteratively with the approximate solution $m\approx -(W_0/B_0)a$. Using this in the expression \pref{cotomegah} we find $\cos\omega_h={\rm O}(a\Lambda_{\rm QCD}^2/\mu_h)$. In case of an infinitely heavy charm quark we have $\mu_h\rightarrow \infty$ and $\cos\omega_h=0$. We can assume this to still hold in good approximation even for a finite $\mu_h$, because the physical charm quark mass is sufficiently heavy such that in the ETMC lattice simulations we have $\mu_h\gg a\Lambda_{\rm QCD}^2$. 

\subsection{Tree-level masses and vertices}
From now on we restrict ourselves to maximal twist in both the heavy and the light sector, i.e.\ $\omega_h =\phi_l=\pi/2$. 
In this case the ground state reduces to $\Sigma_{\rm V}={\rm diag}(i\sigma^3,1)$. Expanding the LO Lagrangian to quadratic order in the pseudoscalar fields the tree-level masses are easily computed. The pion masses reproduce the familiar 2-flavor results \cite{Scorzato:2004da,Sharpe:2004ny,Aoki:2004ta},
\bea
\mpic & =&  2B_0\mu_l\,\label{LOmpic}\\
\mpin &=& \mpic +\dmpi, \quad \dmpi\,=\, 2c_2a^2\,.\label{LOmpin}
\eea
The tree-level result for the four degenerate kaon masses reads
\bea
m^2_K &=& B_0(m_s + \mu_l)  - \frac{2\hat{a}^2}{f^2} \left(2W^{\prime}_{8}-W_B\right)\,.\label{LOmk} 
\eea
Finally, the  eta mass is related to the kaon and pion mass by the Gell-Mann--Okubo relation,
\bea
m^2_{\eta}&=& \frac{1}{3}\Big(4m^2_K - \mpic \Big)\label{LOmeta}\,.
\eea
We emphasize that this relation does not hold for nonmaximal twist angles. In that case additional O($a^2$) contributions  appear on the right hand side of \pref{LOmeta}. 

The tree-level masses enter the propagators that appear in the one-loop calculation. The only difference to the propagators in continuum ChPT \cite{Gasser:1984gg} is the mass splitting between the charged and neutral pions, so one has to keep track of the flavor indices for the pions in loop diagrams.

The relevant vertices are obtained by expanding \pref{LOLagLCE} to four powers in the pseudoscalar fields.
Expressed in terms of the tree-level masses, \pref{LOmpic}$-$\pref{LOmk}, the vertices stemming from ${\cal L}_2$ are the vertices known from continuum ChPT. Additional vertices stem from ${\cal L}_{a^2}$, and the O($a^2$) vertices read
\bea\label{eqn:nlomaxtwist}
 {\cal L}_{a^2,4 \pi}  & =&  \phantom{+}\frac{8}{3} \frac{\hat{a}^2}{f^4} {W}_{68}' \pi_3^2 \pi^2 + \frac{1}{2} \frac{\hat{a}^2}{f^4} {W}_{78}' \pi_h'^4 \nn\\
	&&+ \frac{8}{3} \frac{\hat{a}^2}{f^4} {W}_{78}' \pi_3^2 \pi_8^2 + \frac{4}{\sqrt{3}} \frac{\hat{a}^2}{f^4} {W}_{78}'  \pi_3 \pi_8 \pi_h'^2,
\eea
where we introduced the shorthand notation 
$\pi^2 = \sum_{i=1}^8 \pi_i^2$, $ \pi_h'^2 = \pi_4^2+\pi_5^2-\pi_6^2-\pi_7^2$ and $W^{\prime}_{78}=2W^{\prime}_7 + W^{\prime}_8$.
We emphasize that \pref{eqn:nlomaxtwist} is the result for maximal twist. For arbitrary twist angles many more terms contribute \cite{MAHorz}.

The vertices in \pref{eqn:nlomaxtwist} involve only two combinations of unknown LECs, $W^{\prime}_{68}$ and $W^{\prime}_{78}$. The 2-flavor result in Ref.\ \cite{Bar:2010jk} is correctly reproduced by dropping the heavy kaon and eta field contributions.
Note that the LECs $W_B$ and $W_C$ do not appear explicitly. Their effect is incorporated by expressing the vertices proportional to the quark masses in terms of the pion and kaon masses \pref{LOmpic} - \pref{LOmk}. This is expected, since the two terms in ${\cal L}_{a^2}^{(2)}$ are mass terms and they could be absorbed in a redefinition of the untwisted quark mass. Doing so ${\cal L}_{a^2}^{(2)}$ would not appear explicitly in the chiral Lagrangian.

Recall that the LEC $W_{68}^{\prime}$ is proportional to the tree-level pion mass splitting. Therefore, provided the mass splitting is known from data, the associated four-pion coupling does not involve an unknown LEC. This will play a crucial role later on. 

All the vertices lead to tadpole diagrams that contribute to the various self energies of the pseudoscalars. These diagrams result in standard divergent scalar integrals, which are conveniently regularized by dimensional regularization. The counterterms  necessary for the renormalization are supplied by the NLO Lagrangian $ {\cal L}_{p^2a^2} + {\cal L}_{Ma^2} + {\cal L}_{a^4}$. For the 2-flavor theory this Lagrangian was derived in Ref.\ \cite{Bar:2010jk}. It is straightforward to repeat the derivation for the 3-flavor theory. However, for our purpose here it is not necessary to derive the NLO Lagrangian completely. It is sufficient to derive enough independent terms  that provide the required counterterms for the pseudoscalar masses.

\subsection{One-loop results}

In order to present our results it is useful to follow \cite{Gasser:1984gg} and introduce
\bea
\chLog{P} &=& \frac{m^2_{P}}{32 \pi^2f^2}\log\frac{m^2_{P}}{\mu^2}\,,\quad P\,=\,\pi^{\pm},\pi^0,K,\eta
\eea
as a shorthand notation for the chiral logs. Various combinations of GL coefficients appear and we introduce $L_{46}=2L_6-L_4$ and $L_{58}= 2L_8 - L_5$. With these definitions the NLO results for the charged pion and the kaon mass read
\bea
M_{\pi^\pm}^2 & =& m_{\pi^\pm}^2  \Big [1 +  \chLog{\pi^0} -\frac{1}{3}\chLog{\eta} + 8\frac{ m_{\pi^\pm}^2}{f^2} \left (L_{46} + L_{58}\right ) + 16\frac{m_K^2}{f^2} L_{46} + C_1 \frac{\hat{a}^2}{f^4} \Big ],\label{LCEmpi}\\
M_K^2 & =& m_K^2 \Big [1 + \frac{2}{3} \chLog{\eta} + 8 \frac{m_{\pi^\pm}^2}{f^2} L_{46} + 8 \frac{m_K^2}{f^2} \left (2 L_{46} + L_{58} \right ) \Big ] \nn \\
&&  - \frac{1}{2} \Delta m_\pi^2 \chLog{\pi^0} +  \frac{8\hat{a}^2}{ f^2} {W}_{78}' \chLog{K} + C_2 \frac{\hat{a}^2 m_{\pi^\pm}^2}{f^4} + C_3 \frac{\hat{a}^2 m_K^2}{f^4} + C_4 \frac{\hat{a}^4}{f^6}.\label{LCEmK}
\eea
Here the coefficients $C_i$ are (combinations of) LECs in the NLO Lagrangian. We introduced appropriate inverse powers of $f$ such that these coefficients are dimensionless.\footnote{In that respect our convention differs from the one in Ref.\ \cite{Bar:2010jk} were dimensionful $C_i$'s where introduced.}

A rather trivial check of our results is whether the correct continuum limit is reached. Indeed, for $\hat{a} \rightarrow 0$ we have $m_{\pi^0}\rightarrow m_{\pi^{\pm}}$ and our results reproduce the corresponding ones of continuum ChPT. The charged pion mass has been computed in 2-flavor tmWChPT in \cite{Bar:2010jk}, and \pref{LCEmpi} reproduces this result as well once the contributions from the kaon and the eta are dropped.

In case of the kaon mass the first line in \pref{LCEmK} is again the continuum ChPT result of Gasser and Leutwyler. The second line contains the corrections due to the nonzero lattice spacing. The analytic corrections are expected. For example, away from the continuum limit one expects the GL coefficients to depend on the lattice spacing. Expanding $L_{ij}(\hat{a}^2) = L_{ij}(\hat{a}^2=0) + \Delta_{ij} \hat{a}^2$ the analytic terms in the continuum part generate the contributions proportional to $C_2$ and $C_3$.  The new additional chiral logs involving the neutral pion and the kaon cannot be guessed from the continuum result which contains an eta chiral log only. The new chiral logs stem entirely from the two O$(a^2)$ vertices in the first line of \pref{eqn:nlomaxtwist}. 

For the neutral pion mass we find to NLO
\bea	\label{eqn:lcepi0mass}
			M_{\pi^0}^2 &=& m_{\pi^\pm}^2 \Bigg [1 +  2 \chLog{\pi^\pm} - \chLog{\pi^0} - \frac{1}{3} \chLog{\eta} + 8 \frac{m_{\pi^\pm}^2}{f^2} \left (L_{46} + L_{58} \right ) + 16 \frac{m_K^2}{f^2} L_{46} + \tilde C_1 \frac{\hat{a}^2}{f^4} \Bigg ] \nn\\
			&& + \Delta m_\pi^2 \Bigg [1 - 4 \chLog{\pi^0} - 2 \chLog{K} - \frac{2}{3} \chLog{\eta} + \tilde C_2 \frac{m_K^2}{f^2} + \tilde C_3 \frac{\hat{a}^2}{f^4} \Bigg ] + \frac{32}{3}\frac{\hat{a}^2}{f^2}W_{78}^{\prime} \chLog{\eta}\,.\label{LCEm0} 
			\eea
The coefficients $\tilde{C}_i$ are NLO LECs different from the ones in \pref{LCEmpi} and \pref{LCEmK}. Also this result converges to the correct continuum limit, and it reproduces the result in the 2-flavor theory if we drop all the contributions associated with the kaon and the eta. Taking the difference $ M_{\pi^0}^2-M_{\pi^{\pm}}^2$ we obtain the pion mass splitting to NLO. It has a rather complicated mass dependence with chiral logs involving all pseudoscalars. 

\subsection{Finite volume corrections and a first numerical test}

In deriving our results we assumed an infinite space-time volume. 
Corrections due to a finite spatial volume \cite{Gasser:1987zq} are easily included. The FV corrections essentially amount to a simple replacement of the chiral logarithms, $\mu_P \,\rightarrow \,\mu_P + \delta_{{\rm FV},P}$. Following the notation of Ref.\ \cite{Colangelo:2005gd} the FV correction is given by
\bea
\delta_{{\rm FV},P} &=& \frac{m^2_{P}}{32 \pi^2f^2} \tilde{g}_1(m_P L)\,,
\eea
with $\tilde{g}_1$ containing a sum over modified Bessel functions. The function $\tilde{g}_1$ drops off exponentially for large arguments, so we may expect the dominant source for FV corrections in the kaon mass to be given by the neutral pion contribution. Note that our result makes a definite prediction for these FV corrections provided  the pion mass splitting is known.    

The ETM collaboration has generated lattice data for two different volumes keeping the other parameters fixed \cite{Baron:2010bv}. These data can be used for a first test of our results. For convenience we have summarized the relevant data in Table \ref{table1}. On the two lattices the central values for the kaon mass
differ by 0.8\%. The statistical errors are about 0.1\% and 0.16\%, respectively, so a FV effect is noticeable in the kaon mass.

\begin{table}[t]
\begin{center}
	\begin{tabular}{c c c c c c }
		Ensemble & $a M_{\pi^\pm}$ & $a M_{\pi^0}$ & $aM_K$ & $a f_{\pi}$ & $L / a$ \\
			\hline
			A40.32&0.1415(04)&0.0811(50)&0.25666(23)&0.04802(13)&32 \\
			A40.24&0.1445(06)&0.0694(65)&0.25884(43)&0.04644(25)&24 \\
			\hline
		\end{tabular}
	\end{center}
\caption{Data for pseudoscalar masses and the pion decay constant taken from Refs.\ \cite{Herdoiza:2013sla,Baron:2010bv}. The data for the decay constant is divided by $\sqrt{2}$ in order to account for the different normalization used in \cite{Baron:2010bv} for the decay constant. The data were generated with $\beta=1.9$ corresponding to a lattice spacing $a\approx 0.09$fm. More details can be found in Ref.\ \cite{Baron:2010bv}.}
\label{table1}
\end{table}

The charged pion mass is about 310 MeV on both lattices, while the neutral pion is significantly lighter with $M_{\pi^0}/M_{\pi^{\pm}}\approx 0.48$ and $0.57$, respectively. So the data is in the LCE regime and our results of the last subsection are applicable. With our result \pref{LCEmK} the relative shift of the kaon mass caused by the neutral pion log, $\epsilon_{\rm r, \pi^{0}} = |M_K(L_1) - M_K(L_2)|/M_K(L_2)$ reads
\bea\label{FVshift}
\epsilon_{\rm r, \pi^{0}}&=& \frac{1}{128\pi^2 m^2_K f^2} \left(\Delta m^2_{\pi} m^2_{\pi^0} \tilde{g}_1(m_{\pi^0} L)\Big|_{L=L_1}
- \Delta m^2_{\pi} m^2_{\pi^0} \tilde{g}_1(m_{\pi^0} L)\Big|_{L=L_2}\right)\,.
\eea
In principle the quantities $\Delta m^2_{\pi}$ and $m^2_{\pi^0}$ are independent of the volume and could be taken out of the difference on the right-hand side. In practice, however, we use the measured values for the pseudoscalar masses and the decay constant. The difference is of higher order in the chiral power counting. Still, the data for the neutral pion mass differ noticeably on the two lattices, although the significance of this difference is questionable in view of the large statistical error. We choose to take the measured central values for the two neutral pion masses and compute $\epsilon_{\rm r}$ following \pref{FVshift}. The result for this procedure reads
\bea\label{estpi0}
\epsilon_{\rm r, \pi^{0}} \approx 0.0024(7)
\eea
The error in this estimate is completely dominated by the error for the neutral pion mass. The estimate \pref{estpi0} falls short by a factor 3 in explaining the observed FV effect. Nevertheless, it has the correct order of magnitude. In contrast, the FV shift due to the eta leads to a shift $\epsilon_{\rm r, \eta}\approx 5\cdot 10^{-5}$.\footnote{The eta mass is given by \pref{LOmeta}. Alternatively it can be taken from Ref.\ \cite{Ottnad:2012fv} with no difference on our estimate.} This is about 50 times smaller than the $\pi^{0}$ contribution and cannot explain the measured FV effect. Note that the eta contribution is the only one in both continuum ChPT and in WChPT in the GSM regime at NLO. Higher order corrections from the charged pion are captured in the resummed formulae of Ref.\ \cite{Colangelo:2005gd}. This contribution is estimated to be of the same order but smaller than the neutral pion contribution \pref{estpi0}.\footnote{See figure 5 of Ref.\ \cite{Colangelo:2005gd}.} More data at various volumes and with different pion mass splittings is needed to fully settle the origin of the observed FV correction. Still, it seems safe to conclude that the FV correction due to the neutral pion needs to be included in analyzing the ETMC data.

\section{Concluding remarks}\label{sec:concl}

Current twisted mass lattice QCD simulations show a sizable pion mass splitting due to explicit flavor symmetry breaking. Twisted mass WChPT provides  formulae which can be used to assess the impact of a large pion mass splitting on the chiral extrapolation and on FV corrections caused by small neutral pion masses. In the case of 2-flavor WChPT such formulae were already derived some time ago. 

The extension to 3-flavor WChPT is slightly nontrivial. The reason is the charm quark that forms a twisted mass doublet together with the strange quark. This ties together strange and charm even if the charm quark is too heavy for the $D$ mesons to be described by ChPT. In order to construct the 3-flavor WChPT Lagrangian we first integrated out the charm quark on the level of the Symanzik effective theory. The resulting 3-flavor theory contains more terms in the effective action than a 3-flavor theory without charm. Still, the standard spurion analysis can be applied to this effective action and the 3-flavor chiral Lagrangian can be constructed as usual.

Based on this 3-flavor chiral Lagrangian we computed the pseudo Goldstone boson masses to NLO in the LCE regime. As anticipated, additional chiral logs proportional to $a^2$ show up at this order, leading to a modified quark mass dependence. The final results contain quite a few additional LECs, and it remains to be seen if there are enough data to resolve all the additional terms in chiral fits.

The additional chiral logs imply additional FV corrections, in particular FV corrections from the neutral pion. Since this is by far the lightest pseudoscalar, these FV corrections are the dominant ones. The LECs entering this correction are directly related to the pion mass splitting. Therefore, these FV corrections are a parameter free prediction of our results if the mass splitting is known. 
A first comparison with numerical data showed that these FV corrections are in the ballpark, but cannot explain alone the observed FV effects in the kaon mass. A careful analysis including the higher order FV corrections due to the charged pion is needed to shed light on this issue.
 
A natural next step is the computation of the decay constants in the 3-flavor theory. It requires the expression for the physical axial vector currents, which can be constructed following the steps we used for the construction of the effective Lagrangian. We expect modifications of the chiral formulae analogous to the ones we found for the masses, in particular larger FV corrections caused by neutral pion logs.

It is also interesting to study  scattering processes. Pion-pion scattering was studied in Ref.\ \cite{Aoki:2008gy,Aoki:2008zk}, and it was shown that the $\pi$-$\pi$ scattering length provides a handle to compute the pion mass splitting without the need to compute disconnected diagrams. Looking at the interaction vertices in \pref{eqn:nlomaxtwist} we expect that the  $K$-$K$ scattering length provides a handle on the LEC $W^{\prime}_{78}$. 

We finally remark that the 3-flavor Lagrangian derived here is also the first step for the description of the mixed action simulations of the ETM collaboration. In order to avoid an unwanted mixing in the heavy sector, simulations with Osterwalder-Seiler valence quarks \cite{Osterwalder:1977pc} are performed as described in \cite{Frezzotti:2004wz}. Mixed action ChPT \cite{Bar:2002nr,Bar:2005tu} takes into account the different discretization effects in the valence and sea sector. The chiral Lagrangian for the latter is the one we have derived here.

\section*{Acknowledgments}
We thank the referee for pointing out a flaw in our first derivation of the chiral Lagrangian and for various valuable comments and suggestions for improvements on the manuscript. 

This work is supported in part by the Deutsche Forschungsgemeinschaft (SFB/TR 09). B.\ H.\ is supported by SFI under Grant No.~11/RFP/PHY3218.

\end{document}